\documentclass{optica-article}

\journal{opticajournal} % for journals or Optica Open

\articletype{Research Article}

\usepackage{lineno}
\begin{document}

\title{Ultrastable vacuum-gap Fabry-P\'erot cavities operated in air}

\author{Yifan Liu,\authormark{1,2,*} Naijun Jin,\authormark{3} Dahyeon Lee,\authormark{1,2} Charles McLemore,\authormark{1,2} Takuma Nakamura,\authormark{1,2} Megan Kelleher,\authormark{1,2} Haotian Cheng,\authormark{3} Susan Schima,\authormark{2} Nazanin Hoghooghi,\authormark{2} Scott Diddams,\authormark{1,2,4} Peter Rakich,\authormark{3} and Franklyn Quinlan\authormark{2,4,$\dag$}}

\address{\authormark{1}Department of Physics, University of Colorado Boulder, Boulder, CO 80309, USA\\
\authormark{2}Time and Frequency Division, National Institute of Standards and Technology, Boulder, CO 80305, USA\\
\authormark{3}Department of Applied Physics, Yale University, New Haven, CT 06520, USA\\
\authormark{4}Electrical, Computer \& Energy Engineering, University of Colorado Boulder, Boulder, CO 80309, USA}

\email{\authormark{*}yifan.liu@colorado.edu}
\email{\authormark{$\dag$}franklyn.quinlan@nist.gov}%% email address is required; see note below about the corresponding author designation

% use {asbstract*} to suppress the copyright line. Copyright information will be added in production

\begin{abstract*} 
We demonstrate a vacuum-gap ultrastable optical reference cavity that does not require a vacuum enclosure. Our simple method of optical contact bonding in a vacuum environment allows for cavity operation in air while maintaining vacuum between the cavity mirrors. Vacuum is maintained long term, with no observed degradation in cavity stability for over 1 year after bonding. For a 1550~nm laser stabilized to a 9.7~mL in-vacuum bonded cavity, the measured Allan deviation is $2.4\times 10^{-14}$ at 1~s and its phase noise is thermal-noise-limited from 0.1~Hz to 10~kHz, reaching about -105~dBc/Hz at 10~kHz offset frequency. This represents the highest stability of any oscillator operated without a vacuum enclosure. Furthermore, we demonstrate a 0.5~mL in-vacuum bonded cavity created using microfabricated mirrors and cavity dicing, with phase noise reaching -95~dBc/Hz at 10 kHz offset frequency. By relieving the need for high-vacuum enclosures, we greatly enhance the portability and utility of low noise, compact cavity-stabilized lasers, with applications ranging from environmental sensing to mobile optical clocks to ultralow noise microwave generation. 
\end{abstract*}

%%%%%%%%%%%%%%%%%%%%%%%%%%  body  %%%%%%%%%%%%%%%%%%%%%%%%%%
\section{Introduction}
Ultrastable laser systems play a pivotal role across various technological and scientific domains, including precision timekeeping\cite{Ludlow_atomic_clock_2015, bloom_optical_2014}, precision spectroscopy\cite{Tamm_2000}, photonic microwave generation\cite{fortier_generation_2011, xie2017photonic, Nakamura2020Science}, and gravitational wave detection\cite{adhikari_gravitational_2014}. The conventional approach to constructing an ultrastable laser system involves locking a laser to a highly stable vacuum-gap Fabry-P\'erot (FP) cavity\cite{Drever_PDH}. These rigid FP cavities, typically ranging in length from several centimeters to half a meter, can achieve remarkable stability through careful design and environmental control. Amazingly, proper cavity mounting and isolation results in vacuum-gap FP cavities that reach the stability limit determined by stochastic volumetric fluctuations in the cavity mirrors and high reflection coatings, resulting in room-temperature length fluctuations below $10^{-16}$ m, equivalent to only a fraction of a proton charge radius\cite{pohl_size_2010}. By combining decimeter-long cavities with cryogenic operation, length instability below $10^{-17}$ m has been achieved, corresponding to a laser fractional frequency instability of only $4\times10^{-17}$ \cite{Matei_2017}.

\begin{figure}
\centering\includegraphics[width=10cm]{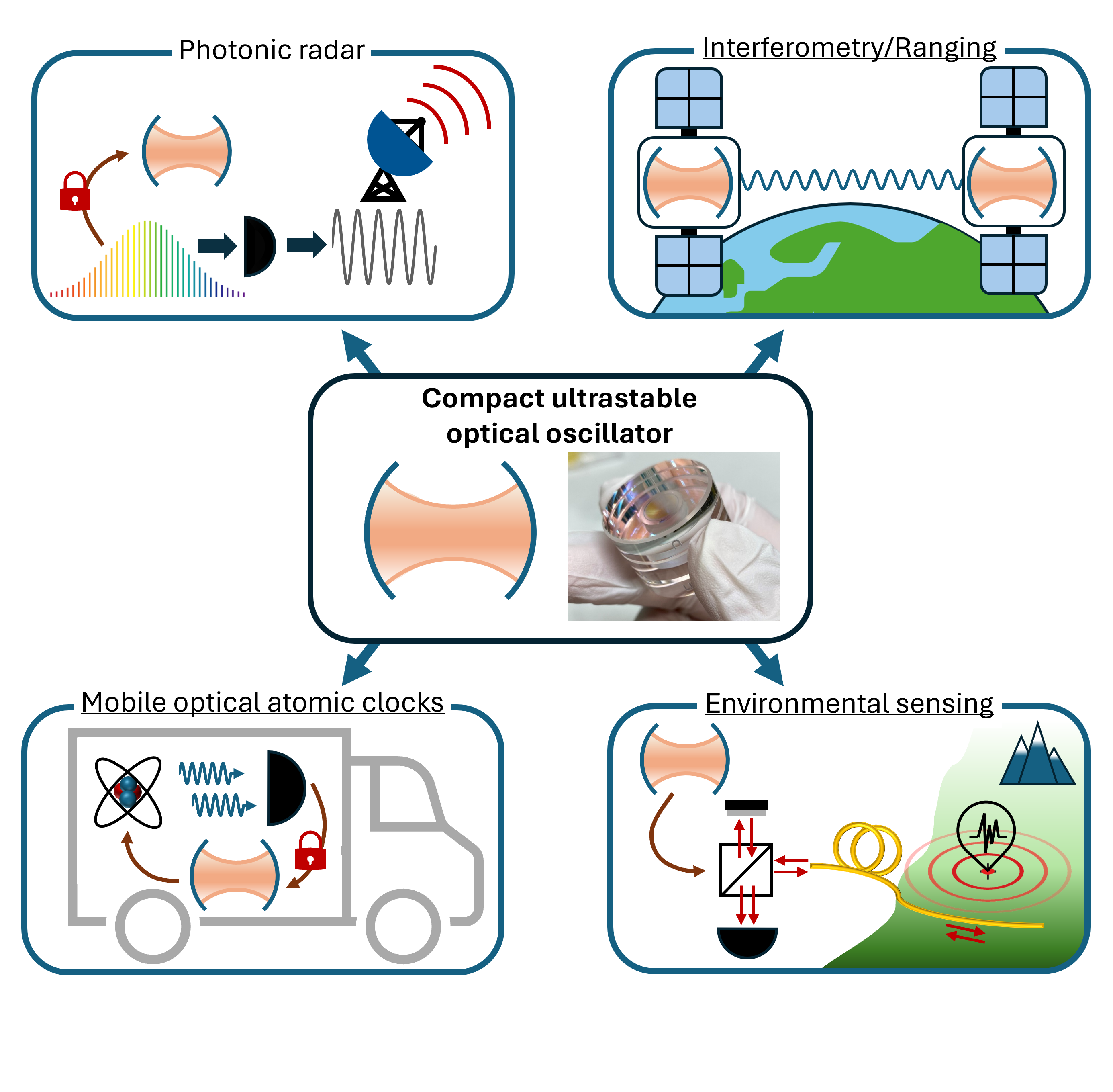}
\caption{\label{Applications}Compact ultrastable laser applications. Compact and field-deployable ultrastable laser systems can be utlized in various applications, such as photonic-based radar systems \cite{Pan_2020,ghelfi_fully_2014}, satellite ranging and interferometry\cite{tapley_grace_2004},  mobile optical atomic clocks\cite{cacciapuoti_space_2009}, and environmental sensing\cite{Marra_2018_Science,Ip_UsingGlobal_2022}.}
\end{figure}

Robust and field-deployable ultrastable laser systems are attractive for out-of-the-lab applications that can benefit from the low phase noise and high stability of ultrastable lasers, such as satellite ranging and interferometry\cite{tapley_grace_2004}, photonic-based radar systems \cite{Pan_2020,ghelfi_fully_2014}, mobile optical atomic clocks\cite{cacciapuoti_space_2009}, and environmental sensing\cite{Marra_2018_Science,Ip_UsingGlobal_2022} (Fig. 1). Traditional vacuum-gap FP cavity systems, despite their superior performance, have significant drawbacks for such out-of-lab applications, mainly due to their large size and weight, as well as the need to maintain ultrahigh vacuum through constant active pumping. Therefore, there have been efforts to miniaturize these laser systems, broadly along two distinct paths: the development of solid-state dielectric resonators and the miniaturization of vacuum-gap FP cavities. Stable dielectric resonators include fiber delay lines\cite{huang_all-fiber-based_2019, jeon_palmsized_2023, Huang_5kmfdl_2023}, integrated spiral resonators\cite{lee_spiral_2013,huang_thermorefractive_2019,Liu_4mcoil_2022}, bulk fused silica FP\cite{stern_ultra-precise_2020,Isichenko_bulkFP_2023, zhang_monolithic_2024}, whispering-gallery-mode (WGM) resonators\cite{zhang_microrod_2019,alnis_thermal-noise-limited_2011}, and stimulated Brillouin scattering (SBS) in ring resonators\cite{loh_operation_2020}. Dielectric resonators, particularly on-chip resonators, can have the advantage of small size and manufacturability with lithographic techniques. But with performance limited by thermorefractive noise and the large thermal expansion coefficient typical of fiber, crystalline, and on-chip waveguide materials, the fractional instability of such resonators is typically above $1\times10^{-13}$. Indeed, it is useful to note that the highest stability dielectric resonators are housed in vacuum enclosures with exquisite temperature control.

Alternatively, the size of traditional vacuum-gap FP cavities can be reduced to make them compact and portable \cite{davila-rodriguez_compact_2017, kelleher_compact_2023, mclemore_miniaturizing_2022}. For example, a vacuum-gap FP with a volume of only 8~mL has been used to demonstrate fractional frequency instability of only $7\times10^{-15}$ at 1 second \cite{ mclemore_miniaturizing_2022}. While such high performance has been demonstrated with greatly reduced cavity size, the requirement for a vacuum enclosure and vacuum pump is not eliminated, and represents a barrier to the realization of compact and portable field-deployable systems.

In this work, we demonstrate laser stabilization with $10^{-14}$ level fractional frequency instability with a sub-10~mL volume cavity while operating without a vacuum enclosure. The cavity performance represents the highest stability ever achieved across optical, microwave or radio frequency domains without vacuum operation. To accomplish this, we have devised a simple and straightforward method to bond the cavity in vacuum, essentially turning the cavity itself into a vacuum cell. With this bonding, and with the cavity surrounded by atmospheric pressure, the optical phase noise is at the mirror coating noise limit across 5 decades of offset frequencies, from 0.1~Hz to 10~kHz, reaching a phase noise level below -100~dBc/Hz at 10~kHz offset. The fractional frequency stability likewise reaches the thermal noise limit, and is only $2.4\times10^{-14}$ at 1 second. For one of our in-vacuum bonded cavities that has been operated for over 1 year, we see no degradation in either phase noise or frequency stability. Furthermore, we combine the in-vacuum bonding method with micro-fabricated mirrors to demonstrate a 0.5~mL-volume miniature cube cavity. This cavity is diced from an array of miniature cavities, and achieves phase noise of -95~dBc/Hz at 10~kHz offset and competitive Allan deviation performance. Thus these in-vacuum bonded cavities enable a significant simplification of the vacuum-gap FP-based ultra-stable laser system, removing the bulky vacuum enclosure and active vacuum pumping requirements, without degrading the performance. With unprecedented frequency stability and phase-noise properties in a compact setup, this work points towards a truly portable and field-deployable ultrastable optical reference.

\section{Vacuum requirement analysis and measurement}
The vacuum requirement for ultrastable FP cavity systems arises from the desire to eliminate cavity length fluctuations coming from variations of the refractive index in the optical beam path, such that the cavity can operate at the thermal noise limit determined by the cavity mirror substrates and coatings. Vacuum in ultrastable FP cavity systems is typically held with an ion pump, since this pump type has no moving parts that could couple vibrations to the cavity. However, ion pumps best operate in the high and ultrahigh vacuum regimes, that is, at a pressure levels below $10^{-6}$~hPa. As we show in this section, this is far below what is required for the cavity to operate at the thermal noise limit.

First, we estimate the residual gas pressure that allows us to reach our cavity thermal noise limit using a model developed for LIGO\cite{zucker_measurement_1994}. In this model, noise caused by refractive index fluctuations is derived from a microscopic picture that considers the effects of individual molecules entering the optical beam path. As shown in more detail in the supplement, assuming a constant beam radius $w_0$ and ideal gas condition, the resulting single-sideband optical phase noise on an optical carrier at frequency $\nu$ at offset frequencies below 100~kHz is well approximated by
\begin{equation}
    \mathscr{L}(f)\approx\sqrt{\frac{m}{2}}\frac{8\pi^2\alpha^2}{Lw_0}\frac{P}{(k_BT)^{3/2}}\frac{\nu^2}{f^2},
\end{equation}
where $\alpha$ is the polarizability of the gas molecules, $m$ is the mass of the individual gas molecule, $L$ is the cavity length, $P$ is the residual gas pressure, $k_B$ is the Boltzmann constant, $T$ is the system temperature, and $f$ is the offset frequency.

A key assumption in the derivation of Eq.~1 is that the residual gas molecules pass through the beam without collisions with other molecules. This means that the mean free path of the molecules should be much larger than the cross-sectional diameter of the beam, which puts an upper bound on the pressure for this assumption to be valid. For our compact optical reference cavities, the average beam radius is around 0.2~mm, such that the applicable pressure range is below about 0.1~hPa (assuming the gas is predominately nitrogen molecules maintained at 300~K). If the cavity length is 6.35~mm and the optical carrier frequency is 1550~nm, Eq.~1 suggests $\mathscr{L}(1~\text{Hz})\approx -44~\text{dBc/Hz}$ and $\mathscr{L}(1~\text{kHz})\approx -104~\text{dBc/Hz}$ at a pressure level of 0.1~hPa, which are well below the thermal noise limit such a cavity can achieve \cite{Kessler:12}. Thus, it should not be necessary to have a high-vacuum environment for the cavity to perform at its thermal noise limit. In fact, the required vacuum level is expected to be quite moderate.

\begin{figure}
\centering\includegraphics[width=12cm]{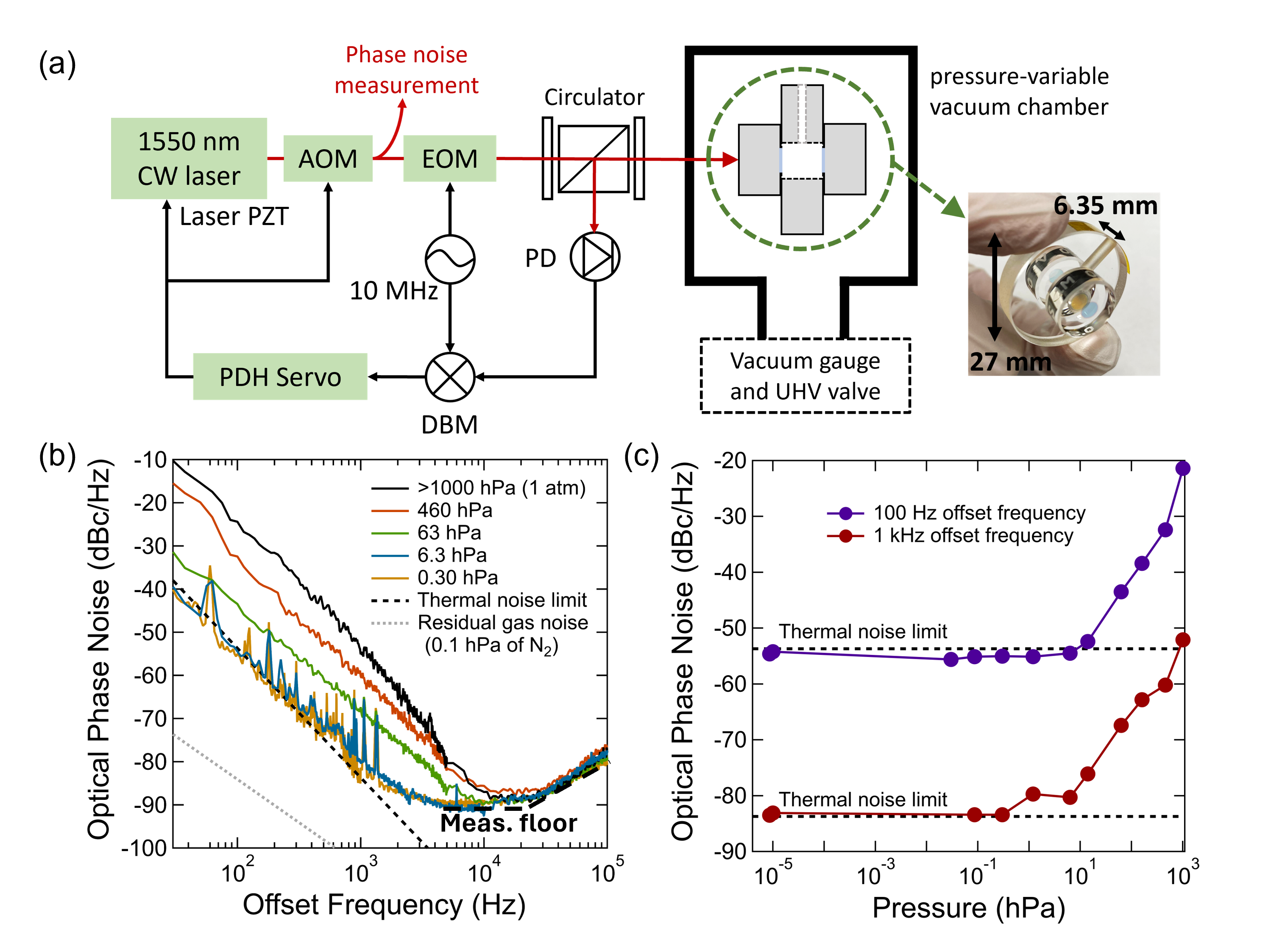}
\caption{\label{vacuum level}Vacuum requirement measurement. (a) Experimental setup. A 1550~nm continuous wave (CW) laser was stabilized via the Pound-Drever-Hall (PDH) method to the 6.35~mm long reference cavity, which was mounted in a pressure-variable vacuum chamber. A portion of the light after the acoustic-optic modulator (AOM) was taken to beat with a stable reference for phase noise measurements at different pressure levels. PZT: piezoelectric transducer. EOM: electro-optic modulator. DBM: double-balanced mixer. PD: photodetector. UHV: ultra-high-vacuum. (b) Selected phase noise traces at different pressure levels plotted with the estimated thermal noise limit of the cavity and the residual gas noise for 0.1~hPa of nitrogen molecules. After 1~kHz offset, the measurement is limited by the signal-to-noise ratio (SNR) of the heterodyne beat. (c) The change of phase noise levels at 100~Hz offset frequency and 1~kHz offset frequency as pressure increases. The uncertainty in the pressure reading is estimated to be about 30\% for the pressures below 100~hPa and 50\% for the pressures above 100~hPa.}
\end{figure}

To verify this conclusion, we experimentally measured the phase noise of a 1550~nm laser locked to a 6.35~mm-long cavity operating at different pressure levels. The experimental setup is shown in Fig.~\ref{vacuum level}a. A 1550 nm commercial fiber laser is locked to the cavity via the Pound-Drever-Hall (PDH) method, with fast feedback control achieved through an acousto-optic modulator (AOM) and slow feedback control through the laser’s piezoelectric tuning port. The cavity has a vent hole and is mounted in a vacuum chamber placed inside an acoustic isolation box on a vibration isolation stage. Connected to the vacuum chamber are a vacuum gauge and an ultra-high-vacuum (UHV) valve. First, we pump the vacuum chamber pressure down to $10^{-7}$~hPa using a turbo pump, then close the UHV valve and disconnect the pump. Without active pumping, the vacuum chamber does not hold at $10^{-7}$~hPa and the pressure will slowly rise. A heterodyne beat is taken between the cavity stabilized light and an optical frequency comb (OFC) fully stabilized to the local oscillator of a Yb atomic clock\cite{schioppo_ultrastable_2017}. While the pressure inside the chamber rises, phase noise measurements are performed on the heterodyne beat at different pressure levels indicated by the vacuum gauge. The experimental results are shown in Fig.~\ref{vacuum level}b and Fig.~\ref{vacuum level}c. The phase noise was limited by the cavity thermal noise limit at low pressures and only starts to deviate from this limit after reaching 0.30~hPa.  (After 1~kHz offset, the measurement is limited by the signal-to-noise ratio of the heterodyne beat and noise from the optical frequency comb.) This measurement confirms that only a very moderate level of vacuum (< 0.3~hPa) is needed for the cavity to be thermal-noise-limited out to an offset frequency of at least 1 kHz, and that continuous vacuum pumping may not be necessary for best phase noise performance.

\section{In-vacuum bonding and cavity performance}
Demonstrations of vacuum-tight glass-to-glass optical contact bonds date back at least 60 years\cite{VANBUEREN1962340}. However, we are unaware of any prior investigation that shows that the achievable vacuum seal from an optical contact bond can support thermal noise-limited, ultrastable cavities, or that the required vacuum level can be held long-term. In this section, we describe experiments where we bond cavities in vacuum, then lock lasers to the cavities without a vacuum enclosure to demonstrate thermal noise-limited phase noise performance. Moreover, with repeated low phase noise measurements in the months since the initial bonding, we establish the longevity of the in-vacuum bond.

\begin{figure}
\centering\includegraphics[width=12cm]{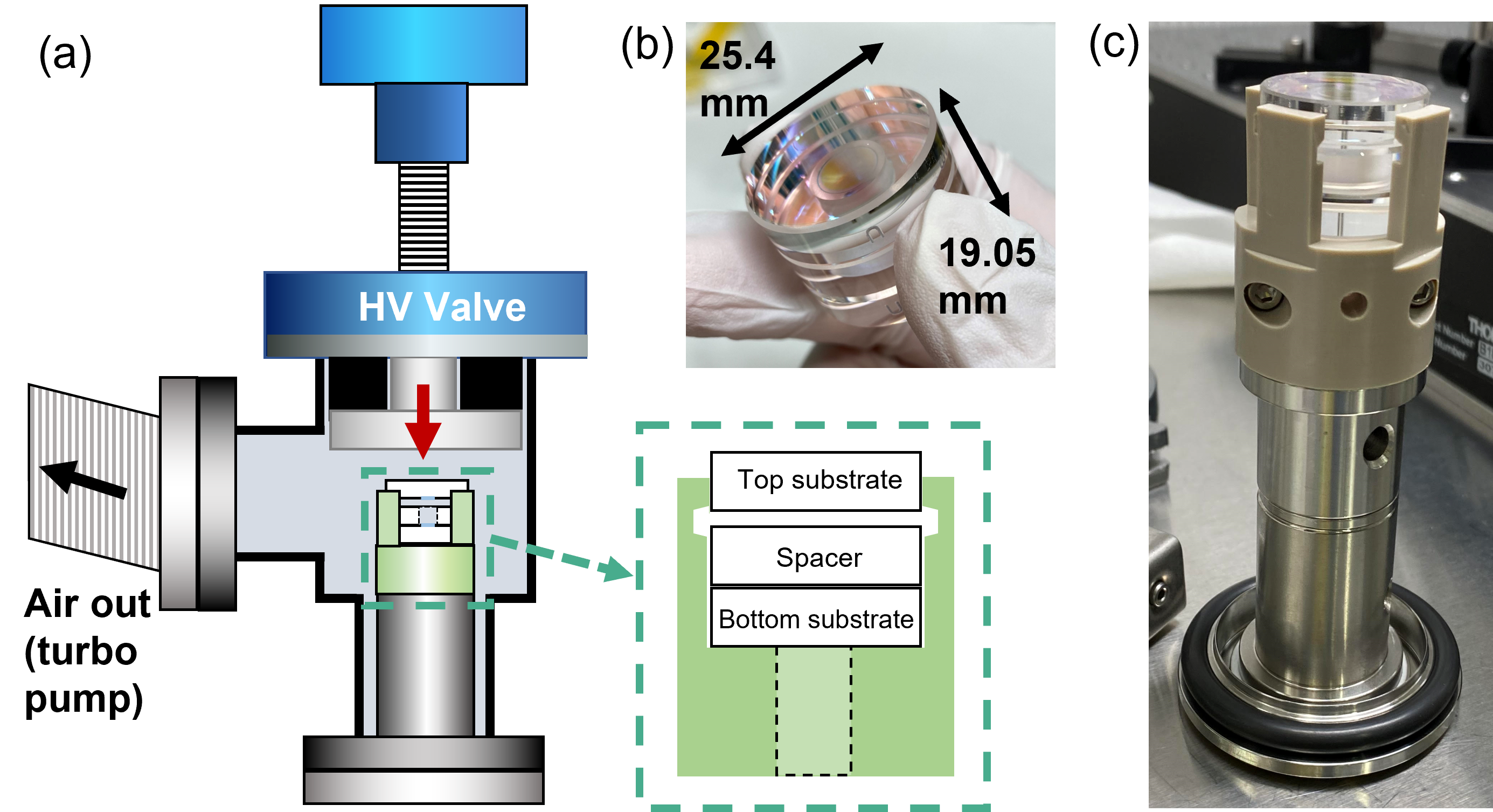}
\caption{\label{Vacuum bonding}In-vacuum bonding method. (a) The mirror substrates and the vent-hole-less spacer are aligned by a vacuum compatible holding structure, which holds the top mirror substrate about 2~mm above the spacer. The holding structure is then placed inside a high-vacuum (HV) bellows valve, which is connected to a turbo pump. Rotating the handwheel of the valve presses the top mirror down and bonds the cavity. (b) Photograph of an in-vacuum bonded cavity. (c) Photograph of the holding structure containing the mirror substrates and spacer.}
\end{figure}

We have bonded several cavities to date using standard off-the-shelf components. Here we describe the essential elements of the bonding technique and results from one of our cavities. More details on the bonding and results from our other cavities may be found in the supplement. The mirror substrates and cavity spacer are all made with ultra-low-expansion (ULE) glass. They have a cylindrical shape with a diameter of 25.4~mm and a length of 6.35~mm. The spacer has a 10~mm diameter center bore hole but, importantly, does not contain a vent hole. One of the mirror substrates has a 1~m radius of curvature (ROC), whereas the other mirror is flat. Both mirrors have a highly reflective (>99.999\%) dielectric coating, and the coating diameter is less than that of the bore hole. The contacting surfaces on the mirrors and spacer are superpolished to the level typical of optical contact bonding, with an average surface roughness of a few angstroms. The in-vacuum bonding setup is illustrated in Fig.~\ref{Vacuum bonding}a. After cleaning, the substrates were placed into a specially designed, vacuum-compatible holding structure that easily aligns the mirrors and spacer concentrically. One mirror substrate was placed at the bottom of the holding structure with the spacer resting directly on top of it. The top mirror substrate is held by friction about 2~mm above the top surface of the spacer. The holding structure containing all three pieces is then placed in a vacuum bellows valve, which is connected to a turbo pump. A vacuum gauge is included between the bellows valve and the turbo pump to monitor the pressure. The pressure quickly falls below $10^{-5}$ hPa, after which we rotate the hand wheel of the valve to lower the inner top surface such that it presses the top mirror substrate into contact with the spacer. While maintaining a compressive force on the bonded cavity, we then shut down and remove the turbo pump. The bonded cavity, surrounded by atmosphere pressure, is then held under the continuous compression force inside the vacuum valve for a few days before being removed for measurements. A picture of an in-vacuum bonded cavity is shown in Fig.~\ref{Vacuum bonding}b, and a picture of the holding structure with mirror substrates is shown in Fig.~\ref{Vacuum bonding}c.

\begin{figure}
\centering\includegraphics[width=13cm]{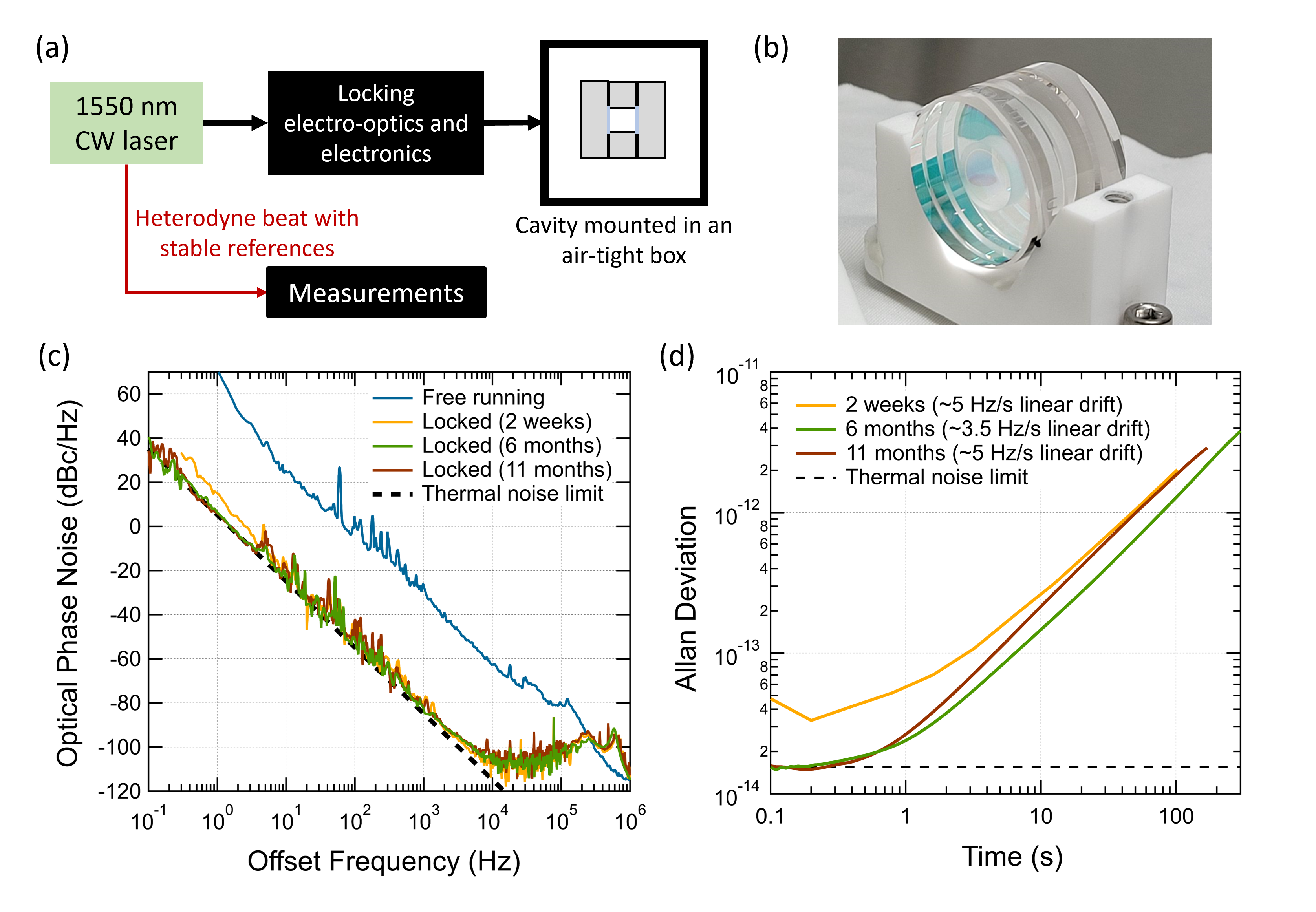}
\caption{\label{Karina}Measurements on the in-vacuum bonded cavity. (a) Simplified diagram illustrating the measurement setup. A 1550~nm fiber laser was stabilized to the in-vacuum bonded cavity via the method of PDH. Part of the stabilized light was taken for phase noise and Allan deviation measurements. The in-vacuum bonded cavity was mounted in an air-tight metal box that is actively maintained at 30$^\circ$C with acoustic and vibration isolation. (b) Photograph of the in-vacuum bonded cavity inside the V-groove holder made with Macor. The cavity rests on two Viton strips. (c) Phase noise of the laser stabilized to the in-vacuum bonded cavity, measured at different times after the cavity was bonded, together with the free-running laser noise and the cavity's estimated thermal noise limit. (d) Allan deviation of the laser stabilized to the in-vacuum bonded cavity, measured at different times after the cavity was bonded, together with the cavity's estimated thermal noise limit.}
\end{figure} 

A 1550 nm laser was locked to the cavity using the PDH technique as shown in Fig.~\ref{Karina}a, with locking electro-optics similar to what is shown in Fig.~\ref{vacuum level}a. The cavity finesse is 591,000, corresponding to a full-width at half-maximum linewidth of 40~kHz. To test the phase noise and frequency stability of the cavity, we mount the cavity into a Macor V-groove holder (Fig.~\ref{Karina}b) enclosed by an air-tight metal enclosure actively maintained at 30$^{\circ}$C. Without vacuum surrounding the cavity, we find an air-tight enclosure critical to keeping the cavity’s temperature and outer-air pressure constant, yielding the best results. Air pressure sensitivity of the cavity is discussed further in the supplement. Since we are primarily concerned with assessing the low noise abilities of in-vacuum bonded cavities, the cavity geometry is not optimized for minimum acceleration sensitivity and was placed within a passive acoustic isolation box with active vibration cancellation.
 
To measure its phase noise, we split the cavity-stabilized light into two channels and compared against two independent ultrastable optical references: an optical frequency comb stabilized to the local oscillator of a Yb optical atomic clock, and a 1550 nm laser stabilized to a 10~cm long vacuum-gap cavity. The two heterodyne beat notes with the two references are digitally sampled and cross-correlated to reveal the noise of our cavity-stabilized laser \cite{davila-rodriguez_compact_2017}. We have performed periodic measurements of the laser phase noise, allowing us to assess the longevity of the vacuum-tight bonding. Representative data are shown in Fig.~\ref{Karina}c along with the cavity's thermal noise limit given by Brownian noise in the coatings. At both 6 months and our most recent measurement at 11 months after bonding, the laser phase noise corresponds to the thermal noise limit of the cavity across nearly 5 decades of offset frequencies, from 0.1 Hz to nearly 10 kHz. At 2 weeks after bonding the noise below 10 Hz offset was slightly elevated, likely due to laser intensity noise coupling to the cavity length \cite{Young_PRL_1999}. Laser intensity noise was improved before the 6- and 11-month measurements were made. As shown in the supplement, our other bonded cavities have been tested with less acoustic isolation and temperature control, and over 1 year since bonding, yet all achieve phase noise limited by the cavity thermal noise, with phase noise $\le-100$ dBc/Hz at 10 kHz offset. 
 
The laser's fractional frequency stability was also recorded, and is shown in Fig.~\ref{Karina}d. For measurements at 6 months and 11 months after bonding, the Allan deviation (ADEV) reaches the thermal noise limit at $1.6\times10^{-14}$ for timescales less than 0.3 second. At 1 second, the ADEV is below $3\times10^{-14}$, and for longer timescales corresponds to a cavity drift rate in the range of 3-5 Hz/s. While the drift rate does vary, it has remained less than or equal to 5~Hz/s. The long-term drift is slightly larger than is typical of all ULE vacuum-gap cavities that are held in thermally shielded vacuum enclosures. Sources of drift are still under investigation. However, we note that without vacuum surrounding the cavity the temperature isolation is reduced, and that our spacer was not acid etched after machining, which is known to improve the dimensional stability of ULE glass\cite{Mattison_1985,JWBerthold_1977}. Acid etch is common practice for ultrastable cavity spacers and will be incorporated in future cavity builds.

\section{In-vacuum bonded cavity with lithographically fabricated mirrors}
The successful creation of ultrastable cavities with in-vacuum bonding creates the opportunity to construct extremely compact cavity-stabilized laser systems. Towards this end, we combined the in-vacuum bonding technique with microfabricated mirrors to create an ultrastable resonator whose volume is only 0.5~mL.

\begin{figure}
\centering\includegraphics[width=13.5cm]{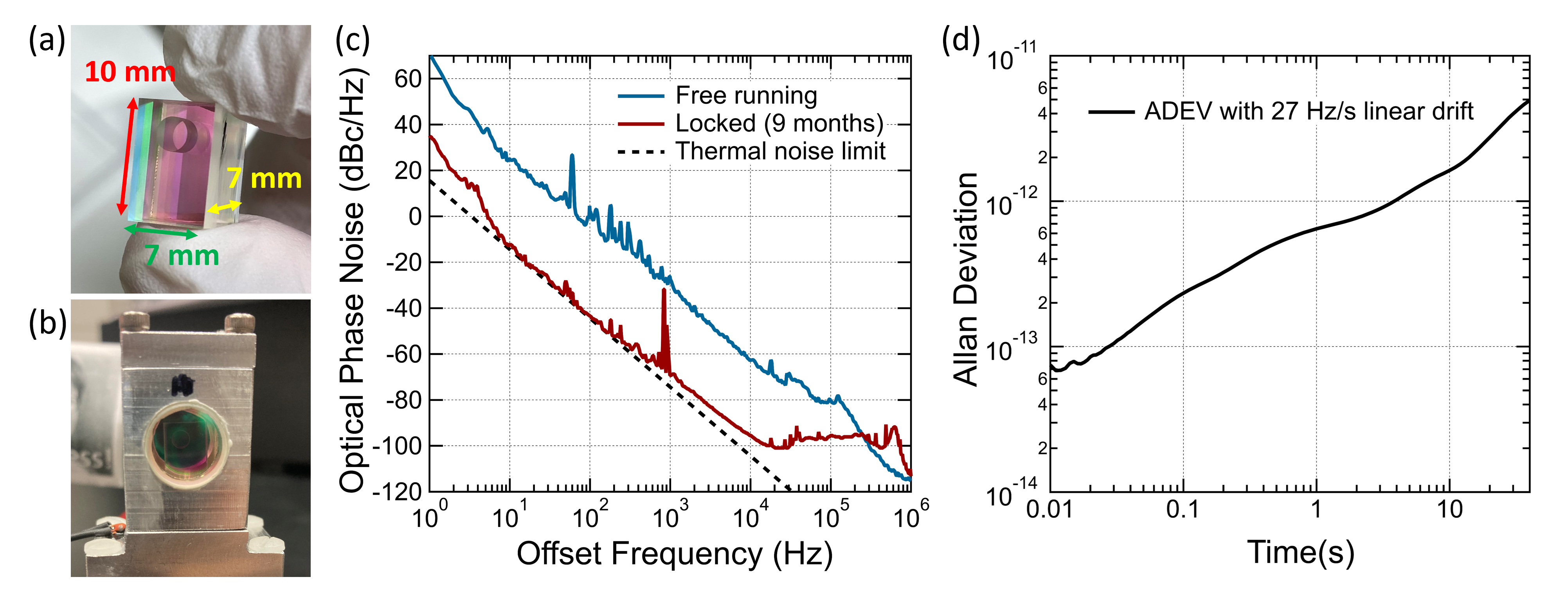}
\caption{\label{minicube}Measurements on the in-vacuum bonded cavity with lithographically fabricated mirror. (a) Photograph of the cavity under test that was diced out of an array of cavities. (b) Photograph of the cavity inside the air-tight metal enclosure. (c) Phase noise measurement on a 1550~nm laser stabilized to the cavity. Blue: free running laser noise. Red: phase noise at 9 months after bonding. Black: estimated thermal noise limit of the cavity. (d) The corresponding measured Allan deviation.}
\end{figure}

We fabricated an array of nine high-finesse, 35-cm-ROC mirrors on a 2-inch-diameter, 2-mm-thick ULE substrate, using the technique described in \cite{Jin_2022}. The mirror substrate was matched to a 2-inch-diameter, 3-mm-thick ULE spacer with nine bore holes, and a high-finesse 2-inch-diameter, 2-mm-thick ULE flat. The mirror substrates and spacer were bonded in vacuum, similarly to the procedure described above. After bonding, the multi-cavity structure was annealed and diced into separate individual cavities. More details on the creation of these cavities will appear in a future publication. One of these miniature cavities was then selected for phase noise and frequency stability testing. The measured cavity finesse is 400,000. The cavity was mounted in a purpose-built, air-tight enclosure made of aluminum as shown in Fig.~\ref{minicube}b, and the whole cavity system is placed within a passive acoustic isolation box on a vibration isolation stage without active temperature stabilization. The experimental results are shown in Fig.~\ref{minicube}c and Fig.~\ref{minicube}d. The measured phase noise is thermal-noise-limited from about 10~Hz to about 300~Hz, reaching about -95~dBc/Hz at 10~kHz offset frequency. The Allan deviation is around $7.5\times10^{-14}$ at 0.01~s and $6.5\times10^{-13}$ at 1~s with a linear frequency drift of around 27~Hz/s. The higher drift rate is assumed to be primarily due to the lack of active temperature control, and we anticipate a much lower frequency drift rate and close-to-carrier phase noise in a well-controlled temperature environment. Additionally, intensity noise on the laser may be contributing to an increase in the phase noise for offset frequencies below 10 Hz, but can be readily reduced with a laser intensity servo. Despite this, the low phase noise can immediately be utilized in applications where short-term phase stability is needed. For example, through the use of optical frequency division \cite{fortier_generation_2011}, this extremely compact cavity can support 10 GHz microwave generation with phase noise below -180 dBc/Hz at 10 kHz offset. With this bond-and-dice approach, we foresee a path towards mass production of such miniature ultrastable optical reference cavities.

\section{Discussion and Conclusion}
\begin{figure}
\centering\includegraphics[width=13cm]{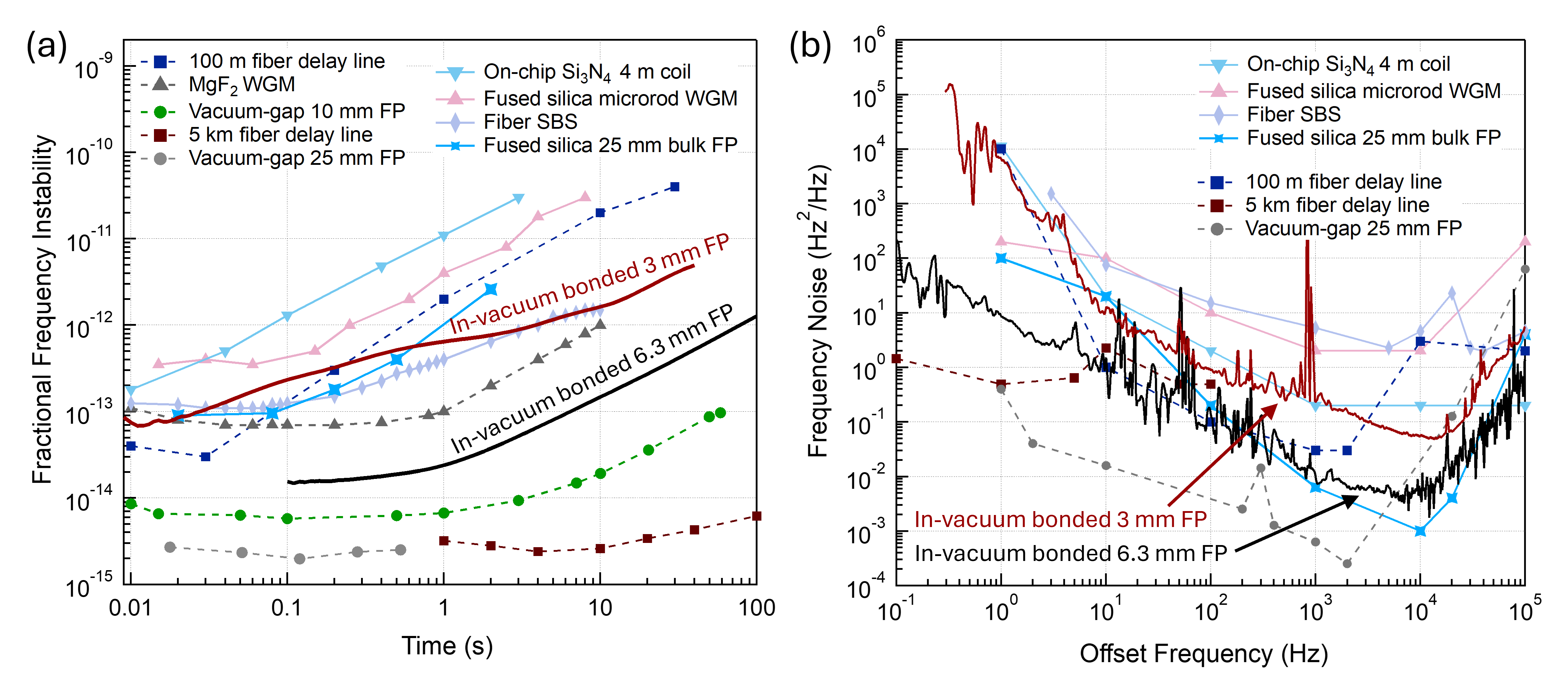}
\caption{\label{Survey}Performance comparison of this work with various state-of-the-art compact optical resonators and ultrastable laser systems. (a) Comparison of fractional frequency instability expressed as Allan deviation. (b) Comparison of laser frequency noise (scaled to 1550~nm). On-chip Si$_3$N$_4$ 4~m coil\cite{Liu_4mcoil_2022}, Fused silica microrod WGM\cite{zhang_microrod_2019}, Fiber SBS\cite{loh_operation_2020}, Fused silica 25~mm bulk FP\cite{zhang_monolithic_2024}, 100~m fiber delay line\cite{jeon_palmsized_2023}, MgF$_2$ WGM\cite{Alnis_2011_WGM}, vacuum-gap 10~mm FP\cite{mclemore_miniaturizing_2022}, 5~km fiber delay line\cite{Huang_5kmfdl_2023}, vacuum-gap 25~mm FP\cite{davila-rodriguez_compact_2017}. Dashed lines represent the presence of vacuum environment.}
\end{figure}

The context by which we may evaluate the performance of our in-vacuum-bonded compact optical reference cavities is shown in Fig.~\ref{Survey}. Vacuum-gap FP cavities are now approaching the form factor associated with on-chip and whispering gallery mode resonators, but with superior phase noise and frequency stability performance. With the optical mode primarily propagating in vacuum, the thermorefractive noise that dominates solid-state dielectric resonators is largely eliminated. Furthermore, the use of ULE can provide orders of magnitude lower thermal expansion coefficient than crystalline, optical fiber, or on-chip SiN resonators, leading to a commensurately lower frequency drift. We note that the lowest noise dielectric resonators require operation in high vacuum and multilayer thermal shielding to achieve their high stability results \cite{alnis_thermal-noise-limited_2011,Huang_5kmfdl_2023}.

Compared to larger vacuum-gap FP resonators, the thermal noise-limited phase noise of our cavities is higher due to the shorter optical cavity length. However, there are many applications for which this higher phase noise floor does not pose a limitation, while a smaller form-factor more easily enables out-of-the-lab use. As noted above, a compact optical reference cavity used in conjunction with an optical frequency comb can support microwave generation with phase noise much lower than conventional electronic sources\cite{fortier_generation_2011}, and common-mode rejection of laser noise in many fiber sensing applications relaxes the phase noise requirements at small offset frequencies\cite{Marra_opticalInterferometry_2022}. Importantly, the higher thermal noise of compact cavities is less of an obstacle at higher offset frequencies – our demonstrated optical phase noise of -105~dBc/Hz at 10~kHz offset is one of the lowest at this offset frequency of which we are aware for a cavity of any size. However, it is also worth noting that, since the phase noise is limited by Brownian noise in the coatings, a longer cavity should result in improved noise performance.

In-vacuum bonded cavities can be further engineered for improved robustness and integration in compact laser systems. While our cavity demonstrations are of simple cylindrical and cubic geometries, in-vacuum bonding is amenable to other designs optimized for low acceleration sensitivity and rigid holding \cite{davila-rodriguez_compact_2017, kelleher_compact_2023, webster_2008, leibrandt_2013, didier_2018, tao2018}. This will be critical for use beyond the laboratory. Furthermore, recent developments in integrating small FP cavities with waveguides on chip \cite{Cheng_2023}, when used in conjunction with low noise on-chip lasers \cite{BohanLi_2021}, should enable a low noise, fully integrated chip-based reference-cavity system. 

In summary, we have developed an in-vacuum bonding technique to provide vacuum-gap ultrastable optical reference cavities that operate surrounded by air. Despite the lack of a vacuum enclosure, the cavities perform at the thermal noise limit determined by Brownian noise of the coatings. By combining in-vacuum bonding with microfabricated mirrors, we created an ultrastable reference cavity whose volume is only 0.5 mL. We anticipate this advance in ultrastable laser technology will accelerate a variety of out-of-the-lab applications, including mobile optical atomic clocks, photonic radar, and sensing.

\begin{backmatter}
\bmsection{Funding}
This work is funded by DARPA and NIST.

\bmsection{Acknowledgments}
We thank Andrew Ludlow and the NIST Yb optical clock team for ultrastable reference light, and Matthew Hummon and Tobias Bothwell
for helpful comments on the manuscript. Product names are given for scientific purposes only and do not represent an endorsement by NIST.

\bmsection{Disclosures}
The authors declare no conflicts of interest.

\bmsection{Data Availability Statement}
Data underlying the results presented in this paper are not publicly available at this time but may be obtained from the authors upon reasonable request.

\bmsection{Supplemental document}
See Supplement 1 for supporting content.

\end{backmatter}

%%%%%%%%%%%%%%%%%%%%%%% References %%%%%%%%%%%%%%%%%%%%%%%%%

%%%%%%%%%% If using BibTeX:
\bibliography{MainText}

\end{document}

\maketitle

\section{Residual gas model}
In the Zucker and Whitcomb model of residual gas noise \cite{zucker_measurement_1994}, a molecule present in the optical beam path will cause a phase shift on the optical field and distort the wave front, leading to a change in the optical path length. The phase shift depends on where the molecule is on the intensity profile of the beam, which can be related to the molecule's velocity, and the net phase shift depends on the total number of the molecules present in the beam path. Summing up the contributions from individual molecules using a Boltzmann distribution of velocity with the assumption of molecules passing through the beam without collisions, the power spectral density of the optical path length fluctuation can be formulated as \cite{zucker_measurement_1994}
\begin{equation}
\label{eqn:PSD}
    S_x(f)=\frac{4\rho(2\pi\alpha)^2}{v_0}\int_0^L\frac{\exp[-2\pi fw(z)/v_0]}{w(z)}dz,
\end{equation}
where $f$ is the offset frequency, $\rho$ is the number density of gas molecules, $\alpha$ is the polarizability of the gas molecules, $v_0=\sqrt{2k_BT/m}$ is the most probable speed of the molecule at temperature $T$ with mass $m$, $L$ is the cavity length, and $w(z)$ is the beam radius. Assuming a constant beam radius $w_0$, converting the length noise into single sideband phase noise on an optical carrier at frequency $\nu$ reduces Eq.~\ref{eqn:PSD} to
\begin{equation}
    L(f)=\frac{2\rho(2\pi\alpha)^2}{v_0Lw_0}\exp(-\frac{2\pi fw_0}{v_0})\frac{\nu^2}{f^2}=\frac{8\pi^2\alpha^2}{v_0Lw_0}\frac{P}{k_BT}\exp(-\frac{2\pi fw_0}{v_0})\frac{\nu^2}{f^2},
\end{equation}
where we have used the ideal gas assumption to express the molecular number density in terms of the pressure and temperature. In the case when $f\ll v_0/(2\pi w_0)$~($\ll$330~kHz for our cylindrical cavity assuming nitrogen molecules), the exponential factor can be ignored. Replacing $v_0$ by $\sqrt{2k_BT/m}$, we arrive at Eq.~1 in the main text.

\section{In-vacuum bonding and additional cavities}
To prepare the ULE components for bonding, the mirror substrates are first cleaned by methanol, rinsed in DI water and dried with dry nitrogen. The spacer is first submerged in a piranha solution at 80$^\circ$C for 10~minutes, rinsed by deionized water, then dried. Except the piranha solution cleaning of the spacer, all the cleaning and bonding processes were done in a laminar flow hood outside the clean room.

\begin{figure}
\centering\includegraphics[width=10cm]{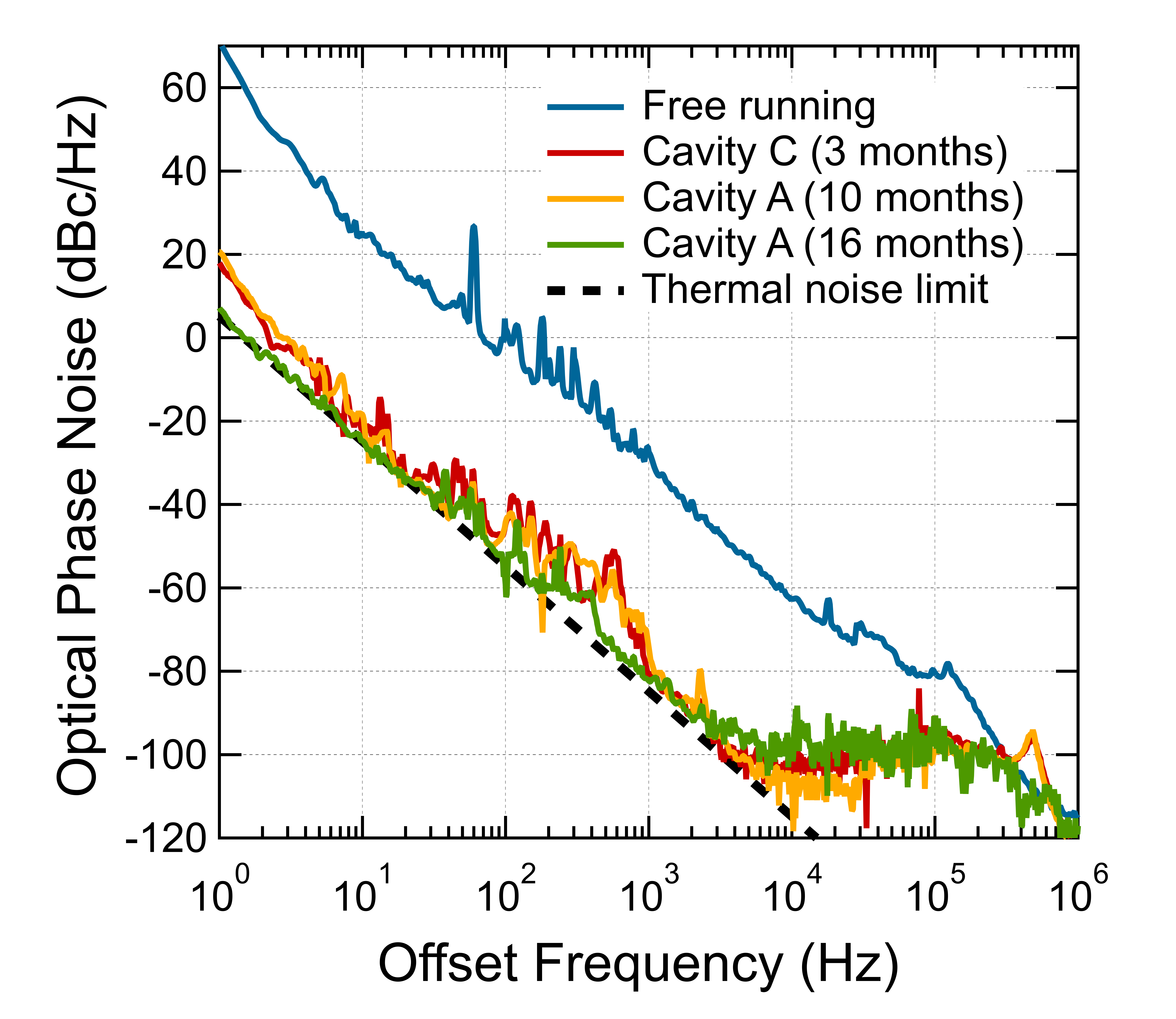}
\caption{\label{AdditionalCavities}Phase noise measurements on a 1550~nm laser stabilized to other in-vacuum bonded cavities, referred to as Cavity A and C, at different times after the cavities were bonded. Also shown is the free running laser noise and the cavities' estimated thermal noise limit. The 16-month measurement was performed in a different system with more optimized settings compared to that of the other two measurements.}
\end{figure} 
 
In some cases, we performed a low-temperature bake-out at about 100$^\circ$C for 4 days to reduce the base system pressure to below $10^{-7}$~hPa. Without bakeout, the pressure during bonding was near $10^{-5}$~hPa. The cavity shown in the main text has undergone the baking process before bonding. However, as shown in Fig.~\ref{AdditionalCavities}, no obvious difference in performance has been observed between the cavities that were baked before bonding and the cavity that was not baked before bonding.

In total, we built and characterized three in-vacuum bonded cavities, all of the same geometry. The measured finesse of the cavities is 853,000 (cavity A), 591,000 (cavity B), and 550,000 (Cavity C), all measured at 1550 nm wavelength, corresponding to linewidths (full width at half-maximum) of 28~kHz, 40~kHz, and 43~kHz. Cavity A was bonded first and without bakeout. Cavity B and C were bonded with bakeout. Cavity B was selected for continuous monitoring of its performance, and the results are presented in the main text. Cavity A and C were not operated continuously, but their performance was still tested occasionally to examine the longevity of the vacuum-tight bonding. The most recent results are shown in Fig.~\ref{AdditionalCavities}. Results from Cavity A measured at 10 months and 16 months after bonding are shown. For Cavity C, the displayed measurement is 3 months after bonding. The 3-month measurement on Cavity C and the 10-month measurement on Cavity A were performed in a much more compact enclosure, with much less acoustic and vibration isolation and no optimization on the laser relative intensity noise (RIN) or residual amplitude modulation (RAM). The 16-month measurement on Cavity A was performed with a system that has been optimized for environmental isolation, as well as RIN and RAM rejection. This measurement is  similar to the data of Cavity B presented in the main text, exhibiting thermal-noise-limited phase noise across a wide range of offset frequencies. The higher noise at 10 kHz for Cavity B is due to a reduced number of cross-correlation averages that resulted in a higher measurement noise floor.

All the phase noise measurements were performed by a commercial phase noise analyzer. The 2-week Allan deviation shown in Fig.~4d of the main text was calculated using a frequency counter on the heterodyne beat between the cavity stabilized laser and a stable optical reference.

All the cavities were held in V-grooves as shown in Fig.~2b. There were no extra thermal shields apart from the air-tight metal enclosure. Also, due to the existence of air within the enclosure, the thermal conductivity of such systems is much higher compared to that of a traditional vacuum-gap cavity operating in a high vacuum chamber. Therefore the cavity in our testing set-up can be more sensitive to the change in the ambient temperature of the lab. 

In addition to temperature, cavity operation in an air-tight enclosure maintains outer pressure stability. Using finite element analysis software, we estimate an outer pressure sensitivity on the length of our cylindrical cavities of about $5\times10^{-11}$/Pa. (For reference, we note that a sound wave with amplitude of 1 Pa would correspond to an intensity of 94 dB.) In our sealed enclosure, this pressure change can arise from a temperature change. Using the ideal gas law, we estimate a temperature-induced pressure change can cause a cavity fractional length change of $2\times10^{-8}$/K. This value is close to the expected coefficient of thermal expansion of the ULE spacer, whose exact value depends on how close the temperature is to the ULE zero-crossing temperature, $T_{zc}$. Thus, we expect direct temperature changes and surrounding pressure changes together to effectively shift $T_{zc}$ to a slightly higher temperature. 

% Bibliography
\bibliography{Supplement}